\documentclass{andp2012}% no class options needed by now
\usepackage[english]{babel}

\usepackage{kantlipsum,cuted}

\usepackage[normalem]{ulem}
\usepackage{amsmath,amssymb}
\usepackage{multirow}
\usepackage{xcolor,soul}
\usepackage{srcltx}
\usepackage{hyperref,graphicx}

\title{Algebraic localization-delocalization phase transition in moving potential wells on a lattice}
\author[S: Longhi]{Stefano Longhi \inst{1,2}\footnote{Corresponding author\quad E-mail:~\textsf{stefano.longhi@polimi.it}}}
\address[1]{Dipartimento di Fisica, Politecnico di Milano,Piazza L. da Vinci 32, I-20133 Milano, Italy}
\address[2]{IFISC (UIB-CSIC), Instituto de Fisica Interdisciplinar y Sistemas Complejos, E-07122 Palma de Mallorca, Spain}
\shortauthors{S. Longhi }
\begin{abstract}
\small
The localization and scattering properties of potential wells or barriers uniformly moving on a lattice are strongly dependent on the drift velocity owing to violation of the Galilean invariance of the discrete Schr\"odinger equation. Here  a type of localization-delocalization phase transition of algebraic type is unravelled, which does not require any kind of disorder and arises when a power-law potential well drifts fast on a lattice. While for an algebraic exponent $\alpha$ lower than the critical value $\alpha_c=1$ dynamical delocalization is observed, for $\alpha> \alpha_c$ asymptotic localization, corresponding to an asymptotic frozen dynamics, is instead realized. At the critical phase transition point $\alpha_=\alpha_c=1$ an oscillatory dynamics is found, corresponding to Bloch oscillations. An experimentally-accessible photonic platform for the observation of the predicted algebraic phase transition, based on light dynamics in synthetic mesh lattices, is suggested.
\end{abstract}
\shortabstract
\begin{document}
\maketitle
% \noindent

\section{Introduction}

In non-relativistic wave mechanics, both space and time are considered continuous and the physical phenomena are invariant under a Galilean transformation,  which is reflected by the covariance of the Schr\"odinger equation for Galilean
boosts \cite{r1}. However, since long time several authors have speculated about the possibility that spatial and/or temporal coordinates can be discrete, introducing variant forms of the Schr\"odinger equation in
which the wave function is defined on discrete lattice sites
of space, time, or space-time, instead of on the spacetime
continuum \cite{r2,r3,r4,r5,r6,r7,r8,r9,r10,r11}. One of the main implications of space and/or time discretization is violation of Galilean covariance of the Schr\"odinger wave equation, leading to the paradoxical result that physical phenomena look different  for different inertial frames of reference. While earlier models of discrete
wave mechanics did not find much attention on a foundational level and space quantization is currently assumed in possible approaches of quantum gravity, space discretization naturally arises when dealing with classical or quantum transport on a lattice, which is described by a discrete Schr\"odinger wave equation \cite{r12,r13,r14,r15,r15b,r16}, as well as in numerical discretization methods to solve the ordinary continuous Schr\"odinger equation \cite{r17,r18}. 
In such lattice systems, violation of Galilean invariance is simply signaled by the non-parabolic nature of the energy-momentum dispersion relation \cite{r19,r20,r21} and is responsible for a variety of intriguing phenomena \cite{r19,r22,r23,r24,r25,r26,r27}, which can be harnessed to control wave scattering, transport and localization in ways which are impossible when space is continuous.
For example, any potential barrier becomes reflectionless or even invisible when sliding fast enough on a lattice \cite{r22,r23,r24}, whereas Airy wave packets on lattice cannot accelerate indefinitely and display relativistic motion \cite{r24}. Interestingly,
Anderson localization of a drifting disorder on a lattice, as well as localization-delocalization phase transitions of  moving incommensurate disorder, are washed out as a result of space discretization \cite{r26,r27}. 

A rather general result rooted in the violation of Galilean invariance is that the bound states of any potential well slowly-drifting on a lattice radiate, i.e. they become resonance states \cite{r19}. Also, in the quasi-continuum limit the number of such resonance states can vary as the drift velocity is varied as a result of a mass renormalization effect \cite{r19}. However, the localization features of drifting potential wells on a lattice in the fast moving regime remain largely unexplored, and an open question is whether dynamical localization is always lost, as for fast sliding random or incommensurate sliding potentials \cite{r26}, or not.\\
In this work we unveil a new type of localization delocalization phase transition of algebraic type  that does not require any kind of disorder and arises when a power-law potential well drifts fastly on a lattice. While for an algebraic exponent $\alpha$ lower than the critical value $\alpha_c=1$ dynamical delocalization is observed, for $\alpha> \alpha_c$ a form of dynamical localization, dubbed asymptotic localization \cite{r28,r29}, is instead realized. At the phase transition point an oscillatory dynamics, related to Bloch oscillations, is found. A photonic platform for the observation of the predicted algebraic phase transition, based on light dynamics in synthetic mesh lattices, is suggested.

\section{Drifting potential well on a lattice}
The starting point of our analysis is provided by the time-dependent discrete Schr\"odinger equation with a drifting potential $V(x, t)=V (x - vt)$, which describes rather generally transport of quantum or classical waves on a
 tight-binding lattice with nearest-neighbor hopping amplitude $\kappa$ and on-site potential $V$ sliding on the lattice with a constant speed $v$ [see Fig.1(a)]. The discrete Schr\"odinger equation  reads \cite{r12,r14,r15,r19,r22,r23,r24}
\begin{equation}
i \frac{d \psi_n}{dt}=- \kappa \left\{ \psi_{n+1}(t)+\psi_{n-1}(t) \right\}+V(n-vt) \psi_n
\end{equation}
 \begin{figure}
  \includegraphics[width=8cm]{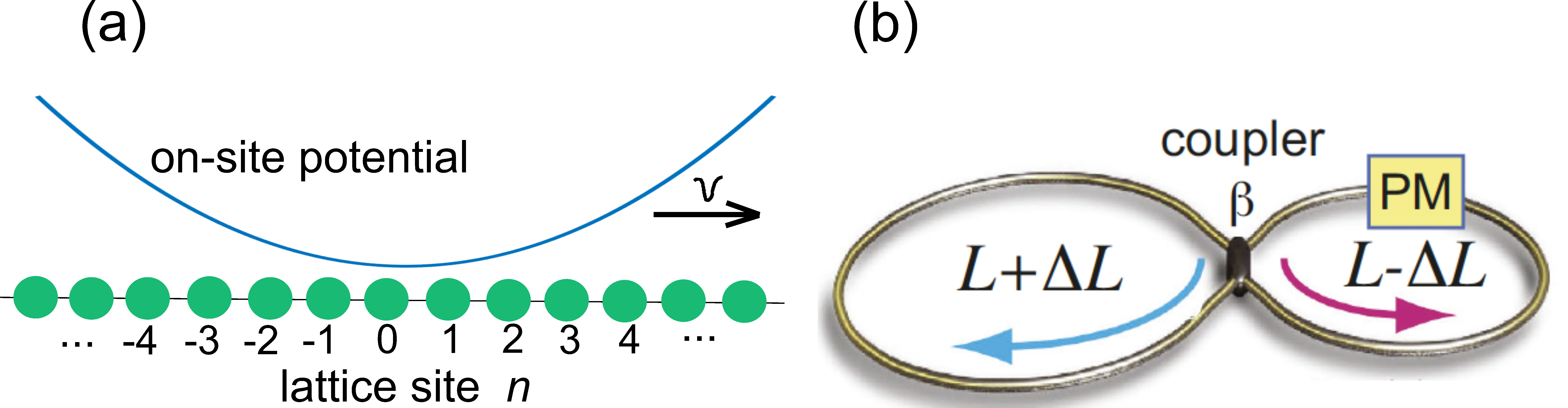}
  \caption{(a) Schematic of a potential well drifting on a one-dimensional lattice at the speed $v$. (b) Physical implementation of the model in a synthetic mesh lattice based on light dynamics in coupled fiber loops with slightly unbalanced lengths $L \pm \Delta L$. A phase modulator (PM), placed in one of the two loops, introduces the sliding on-site potential. The coupling angle between the two fibers is $\beta$.}
 \end{figure}
where $\psi_n(t)$ is the wave amplitude on the $n$-th site of the lattice, $t$ is the continuous time variable, and the space $x$ is scaled to the lattice period. In the following analysis we will focus our attention to an on-site potential well described by a power law, namely 
\begin{equation}
V(x)=V_0 |x|^{\alpha}
\end{equation}
where $V_0$ and $ \alpha$ are positive real numbers. We note that, contrary to the continuous Schr\"odinger equation, for the discrete Schr\"odinger equation the distinction between potential well and potential barrier is 
actually meaningless, since after the gauge transformation $\psi_n (t) \rightarrow (-1)^n \psi_n(t)$ and time-reversal (particle-hole) transformation $\psi_n(t) \rightarrow \psi_n^*(t)$, the sign of $V_0$ can be reversed: this means that the scattering and localization properties of the potential are invariant after flipping the sign of $V_0$, i.e. for potential barriers and wells. For the sake of simplicity, in the following we will continue to refer to a potential well, although the results are valid for a potential barrier as well.

 \begin{figure}
  \includegraphics[width=8cm]{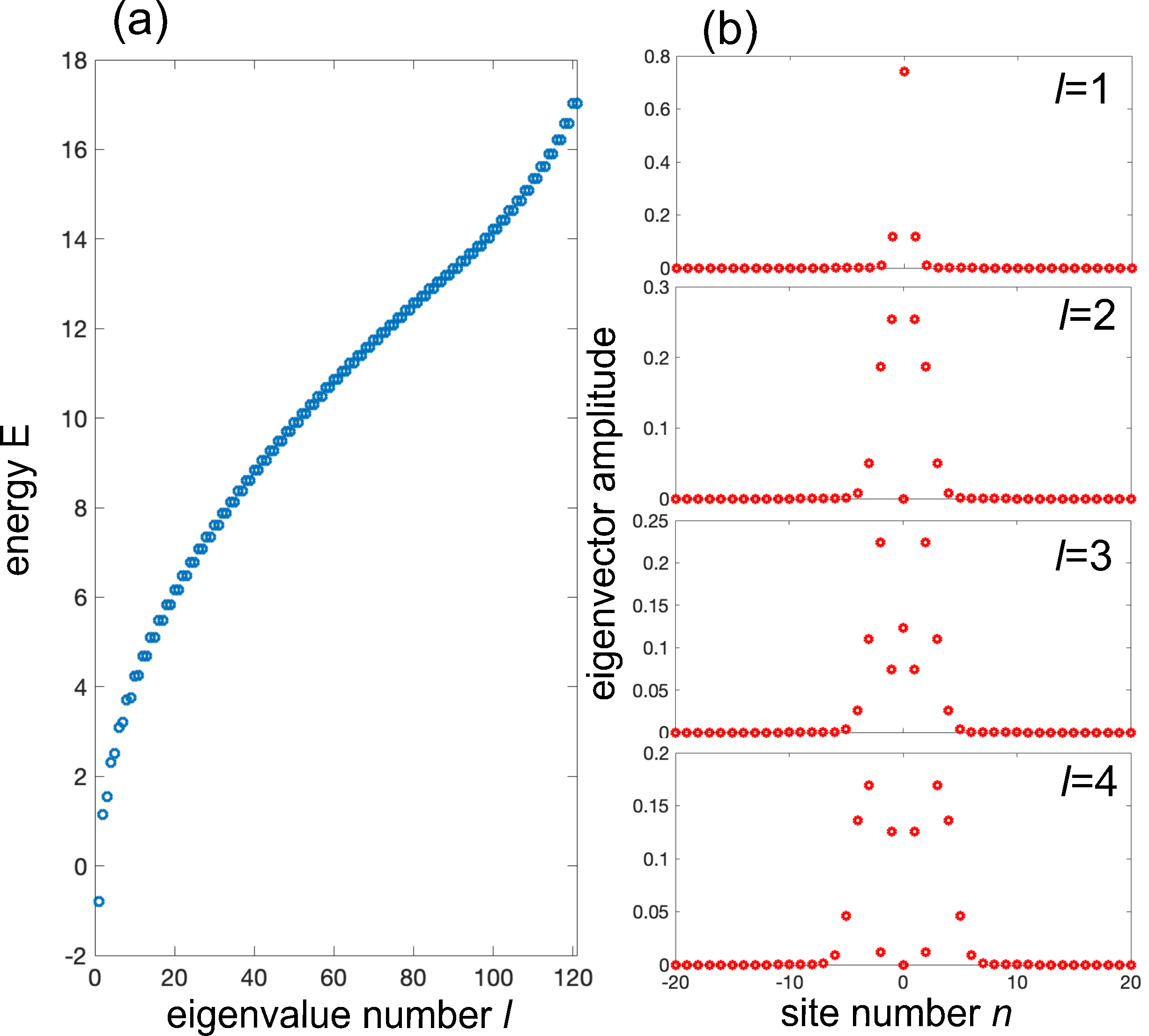}
  \caption{(a) Energy spectrum of the potential well $V(n)=V_0|n|^{\alpha}$  at rest for parameter values $\kappa=1$, $V_0=2$ and $\alpha=1/2$. The energy spectrum has been numerically computed by diagonalization of the matrix Hamiltonian  $H_0$ in a lattice comprising $L=121$ sites, from $n=-60$ to $n=60$, with open boundary conditions. (b) Profiles of the low-order (from $l=1$ to $l=4$) bound states sustained by the potential well (behavior of $|u^{(l)}(n)|^2$). }
 \end{figure}

  \begin{figure*}
  \includegraphics[width=17cm]{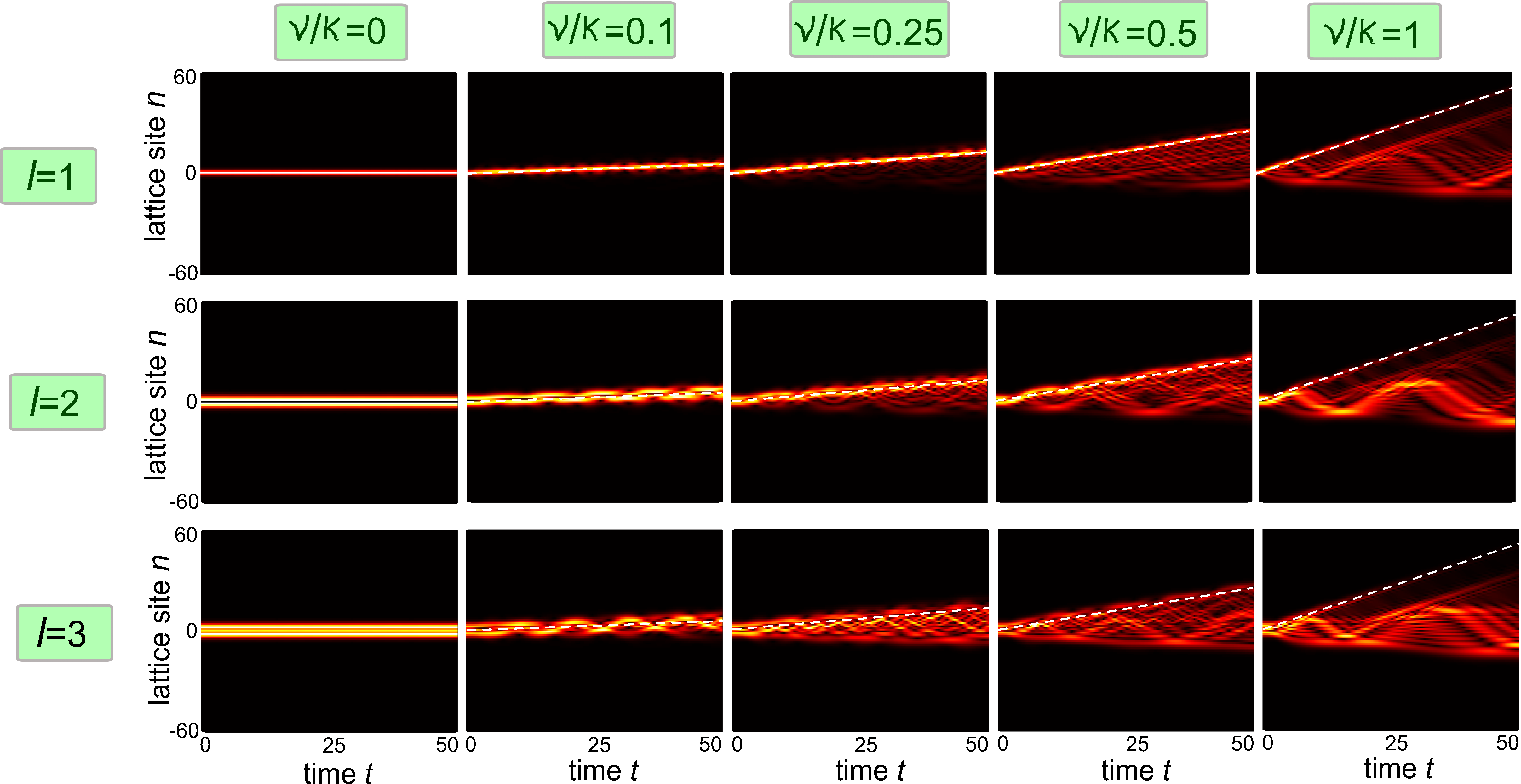}
  \caption{Dragging of low-order bound states of the potential well (2) drifting on the lattice for a few increasing values of the drift velocity $v/ \kappa$.  Parameter values are as in Fig.1 ($\kappa=1$, $V_0=2$, $\alpha= 1/2$). The various panels show the numerically-computed behavior of $|\psi_n(t)|^2$ on a pseudocolor map for a few increasing values of $v/ \kappa$, indicated on the top of the figure. The initial condition $\psi_n(0)=u^{(l)}(n)$ corresponds to the $l$-th bound state of the potential well at rest, with $l=1,2$ and 3 (top, middle and bottom rows, respectively). The straight dashed curves in each panel depict the position of the potential well minimum.}
\end{figure*}

If we consider a moving reference frame $(X,T)$ where the potential is at rest, i.e. after introduction of the Galilean transformation
\begin{equation}
X=n-vt \; , \;\; T=t,
\end{equation}
Eq.(1) takes the form
\begin{eqnarray}
i \frac{\partial \psi}{\partial T} & = & - \kappa \left\{ \psi(X+1,T)+\psi(X-1,T) \right\}+iv \frac{\partial \psi}{\partial X}  \nonumber \\
&+ & V(X) \psi(X,T) \equiv H_v \psi(X,T)
\end{eqnarray}
with the Hamiltonian operator $H_v=-2 \kappa \cos(-i \partial_X)+i v \partial_X+V(X)$.
Unlike for the continuous Schr\"odinger equation, the
drift term $iv(\partial \psi / \partial X)$ on the right-hand side of Eq. (4) cannot
be removed by a gauge transformation \cite{r19}, indicating that Eq.(1) is not covariant for a Galilean boost. Correspondingly, the localization properties of the potential (2) depend on its drift velocity
$v$ on the lattice. 
 Clearly, for the potential well at rest, $v=0$, all eigenstates $u^{(l)}(X)$ of the Hamiltonian $H_0$ are localized and any initially-localized excitation of the system does not spread on the lattice, i.e. dynamical localization is observed. 
Figure 2 shows, as an example, the energy spectrum and the profiles of a few low-order bound states sustained by the potential (2) at rest for $\alpha=1/2$.  
For a slowly drifting potential on the lattice, i.e. for $|v| \ll \kappa$, any bound state of $H_0$ is dragged by the moving potential, as illustrated in the far left panels of Figs.3 and 4. However, the dragging is imperfect, especially for high-order bound states, and leaking is clearly visible as $v$ is increased. For $v$ approaching and above $ \sim \kappa$, the drifting potential is not able anymore to drag the localized bound states, as clearly shown in the far right panels of Figs.3 and 4. The imperfect dragging basically stems from the fact that, in the moving reference frame, the Hamiltonian $H_v$ does not sustain bound states, i.e. the point spectrum of $H_v$ is empty. This means that, for any arbitrarily small drift velocity $v$, any bound state $u(X)$ of $H_0$ becomes a resonance state for $H_v$.  To prove that $H_v$ for any drift velocity $v \neq 0$ does not sustain any bound state, let us suppose by contradiction that $u(X)$ is a bound state of $H_v$ with energy $E$, that is
\begin{equation}
-\kappa u(X+1)-\kappa u(X-1)+i v \frac{du}{dX}+V(X)u(X)=Eu(X)
\end{equation}
with $v \neq 0$. Since by assumption $u(X)$ is a bounded function, with $u(X) \rightarrow 0$ and $V(X) \rightarrow \infty$ as $X \rightarrow \pm \infty$, in the asymptotic limit $X \rightarrow \pm \infty$ the terms $ \kappa u(X \pm 1)$ on the left hand side of Eq.(5) are much smaller than $V(X)u(X)$, and can be thus neglected, yielding
\begin{equation}
i v \frac{du}{dX} \simeq [ E- V(X) ] u(X)
\end{equation}
  \begin{figure*}
  \includegraphics[width=17cm]{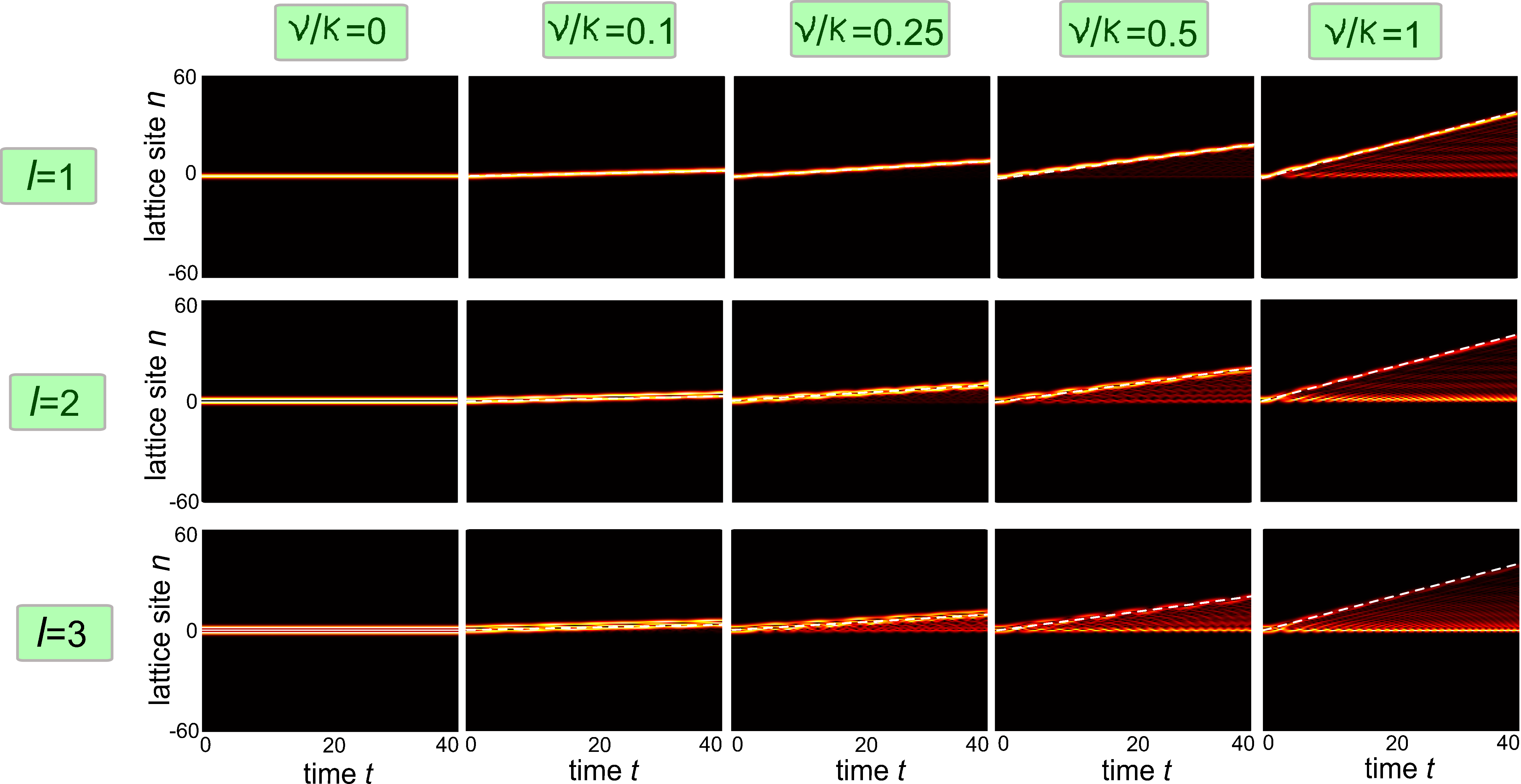}
  \caption{Same as Fig.3, but for a potential well with $\alpha=3/2$ and $V_0=0.5$.}
\end{figure*}
as $X \rightarrow \pm \infty$. Equation (6) can be formally integrated, yielding the following asymptotic behavior of $u(X)$ as $X \rightarrow \pm \infty$
\begin{equation}
u(X) \sim \exp \left\{ -\frac{i}{v} \int^X d \xi \left[E-V(\xi) \right] \right\}.
\end{equation}
Equation (7) indicates that $|u(X)|$ does not decay as $X \rightarrow \pm \infty$, which contradicts the assumption that $u(X)$ is a bound state. Therefore, it not possible for $H_v$ to sustain any bound state when $v \neq 0$. 

 \begin{figure*}
  \includegraphics[width=17cm]{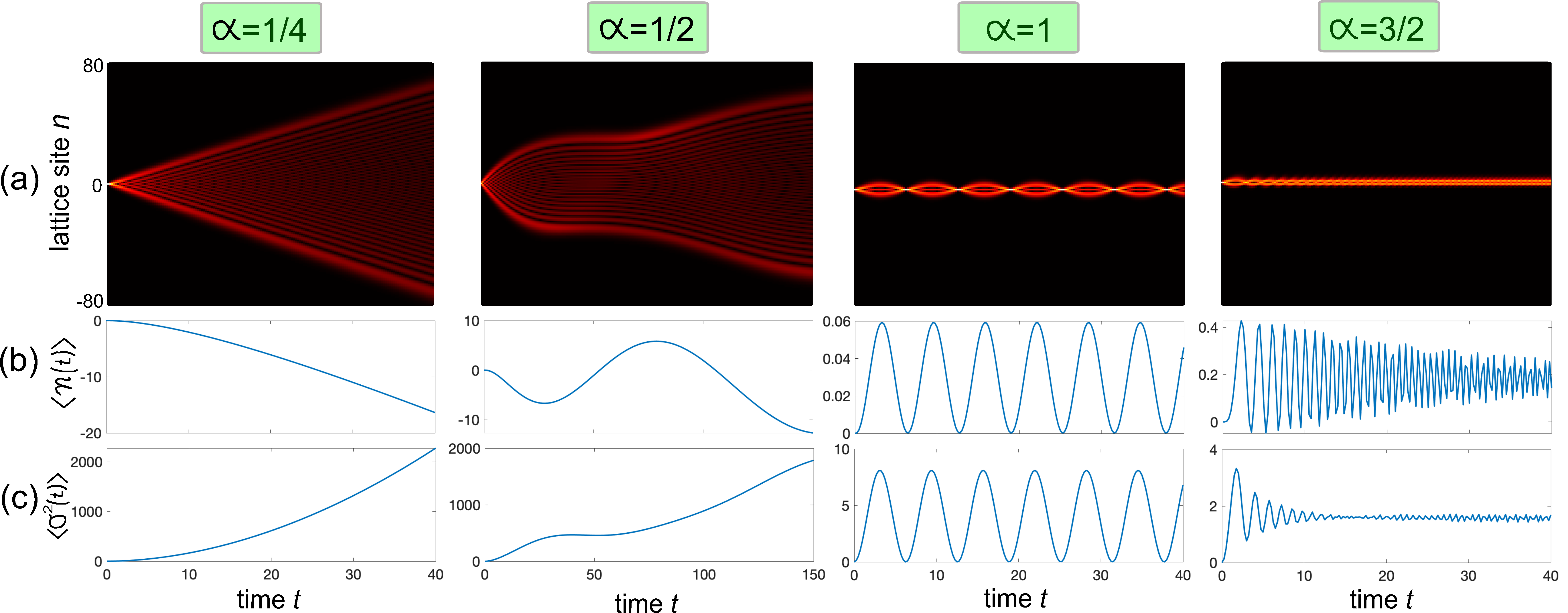}
  \caption{Algebraic delocalization-localization phase transition. The panels show the temporal evolution of (a) the wave function amplitude $| \psi_n(t)|^2$ on a pseudo-color map, (b) wave packet center of mass $\langle n(t) \rangle$, and (c) second moment $\sigma^2(t)$ for initial single-site excitation at $n=0$, for a few increasing values of the power exponent $\alpha$. The hopping amplitude is $\kappa=1$, the drift velocity is $v=5 \kappa$. The potential well amplitude $V_0$ is $V_0=4$ for $\alpha=1/4$,  
  $V_0=2$ for $\alpha=1/2$, $V_0=1$ for $\alpha=1$, and $V_0=0.5$ for $\alpha=3/2$.}
\end{figure*}

\section{Algebraic localization-delocalization phase transition}
Given that the drifting potential well on the lattice cannot drag its bound states and strong radiation is observed in numerical simulations as the drift velocity is increased (see Figs.3 and 4), it is worth investigating the localization properties of the  
drifting potential well in the laboratory $(n,t)$ reference frame in the limit of a fast sliding potential. 
To this aim, let us consider an initial single-site excitation of the system at the site $n=0$, i.e. let us assume $\psi_n(0)=\delta_{n,0}$. The 
dynamical localization properties of the drifting potential are captured by the temporal behavior of the wave packet center of mass $\langle n(t) \rangle$ and second moment $\sigma^2(t)$, defined by
\begin{eqnarray}
\langle n(t) \rangle &  \equiv & \sum_n n |\psi_n(t)|^2 \\
\sigma^2(t) & \equiv & \sum_n (n-\langle n(t) \rangle)^2 | \psi_n(t)|^2.
\end{eqnarray}
Dynamical localization corresponds to a bounded (i.e. not secularly growing) function $\sigma^2(t)$ as $t \rightarrow \infty$. Extended numerical simulations in the fast sliding regime $|v| \gg \kappa$ indicate that a strictly different behavior is observed depending on whether the power exponent $\alpha$ is smaller or larger than the critical value $\alpha_c=1$. Specifically, for $\alpha<\alpha_c$ one observes dynamical delocalization, whereas for $\alpha>\alpha_c$ one has dynamical localization in the form of an asymptotic frozen dynamics. At the boundary $\alpha=\alpha_c=1$ one observes dynamical localization in the form of a periodic (breathing) dynamics, which is characteristic of Bloch oscillation dynamics as discussed below. The main results are illustrated in Fig.5. Contrary to the most common scenario of localization-delocalization phase transitions observed in lattices with random or incommensurate disorder, such as in the Aubry-Andr\'e model where the transition is observed when the potential strength $V_0$ is varied \cite{r30}, in our model the transition is observed when the {\em power exponent} of the potential (rather than its amplitude $V_0$) is varied. For such a reason, the predicted phase transition is dubbed {\it algebraic}.

To understand the dynamical behavior observed in Fig.5 and the appearance of a delocalization-localization phase transition as the power exponent $\alpha$ is varied above the critical value $\alpha_c=1$ in the fast sliding limit, an asymptotic analysis of Eq.(1) can be performed, which is presented in the Appendix A. To get some physical insights, it is worth considering the dynamics in the laboratory $(n,t)$ reference frame and introducing the gauge transformation
\begin{equation}
\psi_n(t)= \phi_n(t) \exp \left( -i \int_0^t d \xi V(n-v \xi) \right),
\end{equation}
so that Eq.(1) takes the form
\begin{equation}
i \frac{d \phi_n}{dt}  =  -\kappa \exp [i \varphi_n(t)] \phi_{n+1}-\kappa \exp[-i \varphi_{n-1}(t)] \phi_{n-1}
\end{equation}
where we have set
\begin{equation}
\varphi_n(t) \equiv \int_{0}^{t} d \xi \left[ V(n-v \xi) -V(n+1-v \xi) \right].
\end{equation}
The form of Eq.(11) allows one for a qualitative asymptotic analysis of excitation transfer between adjacent sites in the lattice, even when the drift velocity $v$ is small.
According to the Landau-Zener theory of avoided level crossing \cite{r31,r32,r33}, excitation transfer between the two sites $n$ and $(n+1)$ occurs around the time instants such that the phase $\varphi_n(t)$ is stationary, i.e. $(d \varphi_n /dt)=0$ or
\begin{equation}
V(n-vt)=V(n+1-vt),
\end{equation}
which from Eq.(2) yields the space-time line
\begin{equation}
t=\frac{n}{v}+\frac{1}{2v}.
\end{equation}
In the neighboring of such space-time line, according to Landau-Zener theory \cite{r31,r32} the transfer probability $P$ that would be obtained if the level crossing were linear would be $P=1-\exp(-2 \pi \Gamma)$, where
\begin{equation}
\Gamma= \frac{\kappa^2}{v \left| \frac{dV}{dx}(n+1-vt)- \frac{dV}{dx}(n-vt) \right|_{t \rightarrow n/v+1/(2v)}}.
\end{equation}
Clearly, in our model for $\alpha<1$ one has $\Gamma \rightarrow \infty$, indicating that transfer is allowed, whereas for $\alpha>1$ one has $\Gamma \rightarrow 0$, indicating that the transfer is forbidden. 
It should be mentioned that in our case the level crossing is not linear except for $\alpha = 1$, and thus the simplest form of $\Gamma$ given by Eq.(15) cannot be used, requiring a more complex analysis involving non-linear level crossing and based on saddle-point methods \cite{r34,r35}. However,
such a simple argument suggests one that excitation transfer in the lattice should strongly depend on the value of the power exponent $\alpha$ being larger or smaller than the critical value $\alpha_c=1$. To get deeper qualitative insights without resorting to complicated saddle-point methods, for our purposes it is enough to estimate the time range for which effective excitation transfer between adjacent sites is allowed. Such a time range is obtained by imposing that the energy offset $ \Delta E(n-vt)=|V(n+1-vt)-V(n-vt)|$ between the adjacent sites, i.e. $| d \varphi_n / dt|$, be smaller than the hopping amplitude $\sim \kappa$, i.e. 
\begin{equation}
\Delta E(n-vt)=V_0 \left| |n+1-vt|^{\alpha}- |n-vt|^{\alpha} \right| < \sim \kappa.
\end{equation}
In fact, when condition (16) is violated the two levels $n$ and $(n+1)$ are far from resonance for excitation transfer to be allowed.
Condition (16) can be solved geometrically (see Fig.6). A completely different behavior is found for $\alpha<1$ and $\alpha>1$, as shown in Fig.6. In the former case [Fig.6(a)] the condition (16) is satisfied for any space-time point $(n,t)$ with $|n-vt|$ not too close to zero, and for any fixed lattice site $n$ and $t \rightarrow \infty$ one has $\Delta E \rightarrow 0$, indicating that excitation transfer is permitted and behaves asymptotically as in a potential-free lattice. This corresponds to the delocalization regime. Conversely,  for $\alpha>1$[Fig.6(b)]  the condition (16) is satisfied in a narrow space-time interval $|n-vt|$ close to zero, and for any fixed lattice site $n$ and $t \rightarrow \infty$ one has $\Delta E \rightarrow \infty$, indicating that excitation transfer is prohibited apart along the space-time line $n=vt$. In the fast moving regime, since the time interval where excitation is allowed scales as $\sim 1/v$, excitation transfer along the line $n=vt$ becomes negligible, and thus the dynamics is basically frozen, a regime similar to so-called asymptotic localization \cite{r29,r30}. We mention that the fast drifting limit is essential to observe asymptotic localization. In fact, in the low drifting regime excitation transfer along the line $n=vt$ is allowed, resulting in complex wave dynamics with excitation spreading between $n=0$ and $n=vt$, as shown for example in the far right panels of Fig.4. An asymptotic analysis of the wave dynamics in the fast moving limit, which provides analytical results of wave spreading and an approximate form of $\sigma^2(t)$, is presented in the Appendix A. The analysis clearly demonstrates the localization-delocalization phase transition.  In particular, in the delocalization phase $\alpha< \alpha_c$ the analysis indicates that the spreading law $\sigma^2(t)$ sensitively depends on the value of the power exponent $\alpha$, approaching a parabolic law, corresponding to ballistic spreading, as $\alpha \rightarrow 0^+$, and to an oscillatory behavior as $\alpha \rightarrow 1^-$, i.e. as the critical point is approached. The critical case $\alpha=\alpha_c=1$ corresponds to a breathing (oscillatory) dynamics (see Fig.5, third column from the left) which is related to Bloch oscillations. In fact, for $\alpha=1$ and considering the space-time dynamics in the half plane $n< vt$ (a similar analysis could be done for space-time dynamics in the other half plane, i.e. for $n>vt$), Eq.(1) reads explicitly
\begin{equation}
i \frac{d \psi_n}{dt}=- \kappa(\psi_{n+1}+\psi_{n-1})+(vt-n) V_0 \psi_n.
\end{equation}
The time dependence in Eq.(17) can be eliminated by the gauge transformation
\begin{equation}
\psi_n(t)= \phi_n(t) \exp \left(-\frac{1}{2}i v V_0 t^2 \right)
\end{equation}
yielding
\begin{equation}
i \frac{d \phi_n}{dt}=- \kappa(\phi_{n+1}+\phi_{n-1})- V_0 n \phi_n,
\end{equation}
which is equivalent to Eq.(17) with $v=0$.
This shows that for $\alpha=1$ (and only for such a special value of $\alpha$) the discrete Sch\"odinger equation remains invariant for a Galilean boost. 
The above equation (19) describes the dynamics of a quantum particle on a tight-binding lattice under a dc force $F=V_0$ which yields the famous Bloch oscillations, i.e. a periodic (oscillatory) dynamics in time with period $ 2 \pi/F$ (see e.g. \cite{r36}).
Therefore the critical case $\alpha=\alpha_c=1$ corresponds to localization of Bloch oscillation type, i.e. yielding an oscillatory-like dynamics, and separates the regimes of delocalization for $\alpha< \alpha_c$ and asymptotic localization, i.e. frozen dynamics, for $\alpha>\alpha_c$.

  \begin{figure}
  \includegraphics[width=8cm]{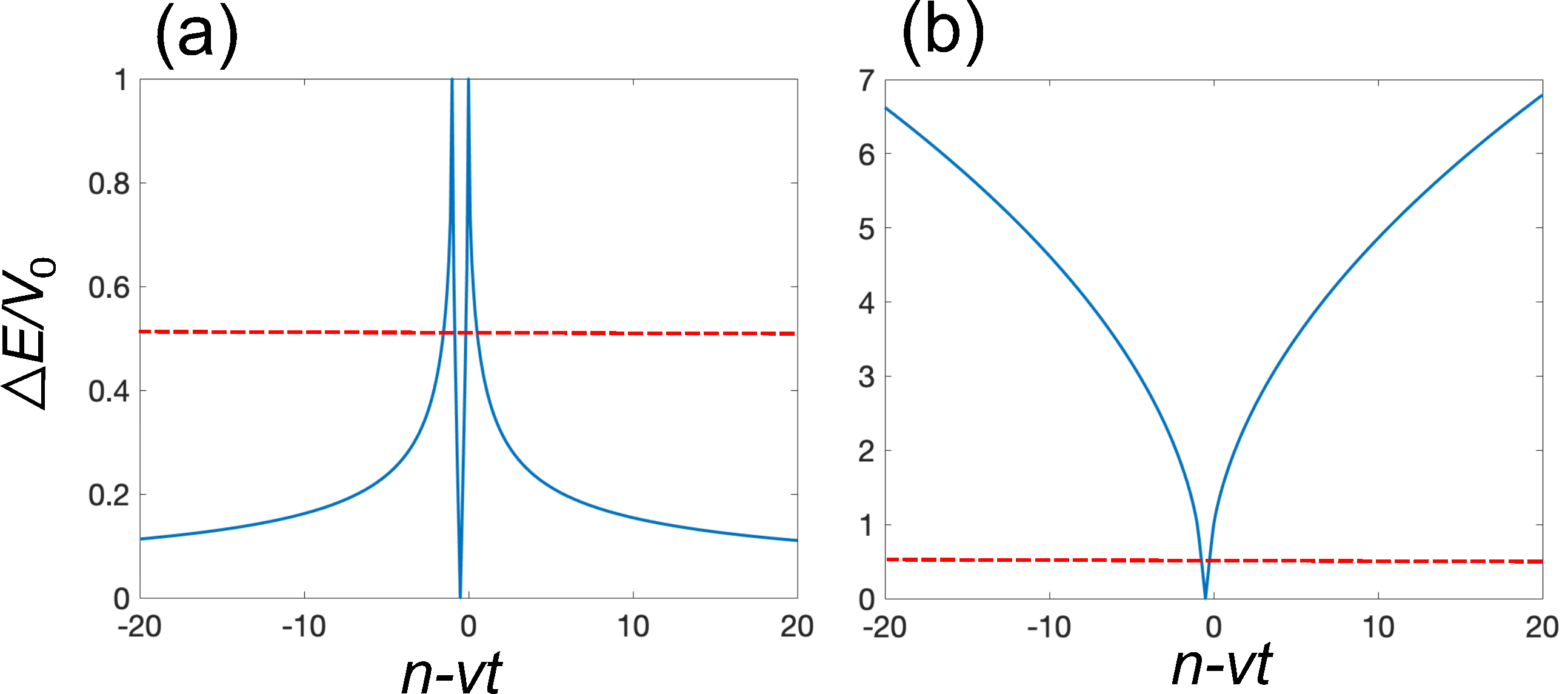}
  \caption{Behavior of normalized energy offset $\Delta E / V_0$ between two adjacent sites of the lattice, versus $(n-vt)$ for (a) $\alpha=1/2$, and (b) $\alpha=3/2$. The dashed horizontal curve gives the reference level $\kappa /V_0$ ($\kappa /V_0=0.5$ in the figure). Excitation transfer between adjacent sites in the lattice is approximately allowed at the space-time points $(n,t)$ such that the curve $\Delta E / V_0$ is below the dashed horizontal curve.}
\end{figure}

\section{Experimental proposal}
Here, we propose a possible experimental platform in photonics for the observation of the algebraic phase transition predicted in the previous section. A main practical difficulty in experiments using standard waveguide lattices fabricated on a sample \cite{r15,r15b,r16,r29} is the possibility to realize sliding potentials with a controllable speed, which is challenging because dynamical changes of the waveguide properties are unfeasible. Such a difficulty can be overcome by considering instead lattices in synthetic space \cite{r37,r38,r39}. A possible setting is provided by light pulse dynamics in mesh optical lattices \cite{r40,r41,r41b,r42,r43,r44,r45} realized by two fiber loops of slightly different lengths $L \pm \Delta L$ connected by a fiber coupler with a coupling angle $\beta$, as schematically shown in Fig.1(b). A phase modulator is placed in one of the two loops, which impresses a phase term to the circulating pulses effectively emulating a drifting potential well. 
Light evolution in the synthetic mesh lattice is described by the set of discrete-time coupled-mode equations  (see e.g. \cite{r40,r41,r41b,r42})
 \begin{eqnarray}
 u^{(m+1)}_n & = & \left(   \cos \beta u^{(m)}_{n+1}+i \sin \beta w^{(m)}_{n+1}  \right)  \exp (-2i\phi_n^{(m)}) \; \\
 w^{(m+1)}_n & = &   \cos \beta w^{(m)}_{n-1}+i \sin \beta u^{(m)}_{n-1}  
 \end{eqnarray}
 where $u_n^{(m)}$ and $w_n^{(m)}$ are the light pulse amplitudes at discrete time step $m$ and lattice site $n$ in the two fiber loops, and $2 \phi_n^{(m)}$ is the phase term impressed to the traveling pulses by the phase modulator at time step $m$. 
 To realize a drifting potential on the lattice at the speed $v$, the phase modulator is driven so that the impressed phase $\phi_n^{(m)}$ is of the form
 \begin{equation}
 \phi_n^{(m)}=V(n-vm)=V_0|n-vm|^{\alpha}.
 \end{equation}
 with $V_0, v \ll 1$. Note that, since we assume $v \ll 1$, the potential $\phi_n^{(m)}$ varies slowly at successive time steps $m$. 
 The discrete-time model described by Eqs.(20) and (21) can be mapped into two decoupled continuous-time discrete Schr\"odinger equations when the coupling angle $\beta$ is close to $\pi/2$ \cite{r46,r47}, thus realizing a sliding potential well on a lattice investigated in the previous sections. In fact,  assuming $\beta=\pi/2-\epsilon$ with $| \epsilon | \ll 1$, 
at first order in $\epsilon$ Eqs.(20) and (21) take the form
  \begin{eqnarray}
 u^{(m+1)}_n & = & \left[   \epsilon u^{(m)}_{n+1}+i  w^{(m)}_{n+1}  \right]  \exp (-2i \phi_n^
{(m)}) \\
 w^{(m+1)}_n & = &   \epsilon w^{(m)}_{n-1}+i u^{(m)}_{n-1}.
 \end{eqnarray}
 From the above equations, one can eliminate from the dynamics the variables $w_n^{(m)}$, yielding a second-order difference equation for $u_n^{(m)}$
 \begin{equation}
 u_{n}^{(m+1)} + u_{n}^{(m-1)} =\epsilon \left( u_{n+1}^{(m)} +  u_{n-1}^{(m) }\right) - 2 i \phi_n^{(m)} u_{n}^{(m+1)}
 \end{equation}
 where we assumed $|\phi_n^{(m)}| \sim \epsilon$. The above equation can be solved by letting 
 \begin{equation}
 u_n^{(m)}=(\pm i)^m \psi^{\pm}_n(m)
 \end{equation} 
 where the amplitudes $\psi_n^{\pm}(m)$ vary slowly with respect to $m$ and satisfy the decoupled continuous-time Schr\"odinger equations
 \begin{equation}
 i \frac{d \psi_n^{\pm}}{dm}= \pm \frac{\epsilon}{2} (\psi_{n+1}^{\pm}+\psi_{n-1}^{\pm})+V(n-vm) \psi_n^{\pm}
 \end{equation}
 which coincide with Eq.(1) after the formal substitution $m \rightarrow t$ and $\epsilon \rightarrow \mp 2 \kappa$.
 
   \begin{figure}
  \includegraphics[width=8.5cm]{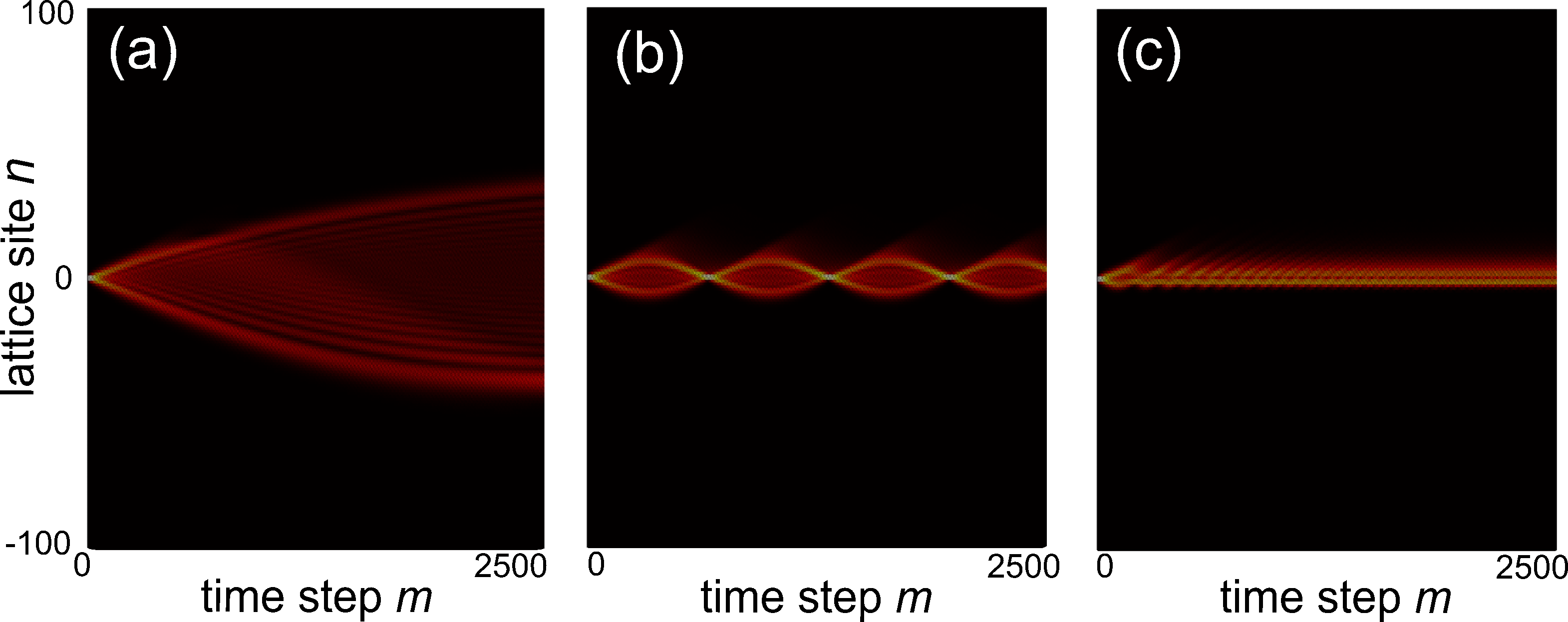}
  \caption{Localization-delocalization phase transition in the synthetic mech lattice system of Fig.1(b). The panels illustrate the evolution at successive discrete time steps $m$ of the pulse light intensity $|u_n^{(m)}|^2+|w_n^{(m)}|^2$ in each time slot (lattice site) $n$ of the synthetic lattice on a pseudo-color map. In (a) $\alpha=1/2$ (delocalization regime), in (b) $\alpha=\alpha_c=1$ (Bloch oscillation regime), in (c) $\alpha=3/2$ (asymptotic localization regime). Other parameter values are given in the text.}
\end{figure}
 
 An example of algebraic localization-delocalization transition in the synthetic mesh lattice system is shown in Fig.7. The figure illustrates the numerically-computed light evolution in the 
 coupled fiber loops as obtained by numerical simulations of the discrete-time wave equations (20) and (21) for a coupling angle $\beta= 0.98 \times \pi/2$, corresponding to a coupling constant $\kappa = \epsilon /2 \simeq (1/2) \cos (\beta)\simeq  0.0157$, and for potential wells drifting on the lattice at the speed $v=2.5 \kappa=0.0393$. Initial condition corresponds to single pulse excitation of the system, namely $u_n^{(0)}=\delta_{n,0}$ and $w_n^{(0)}=0$. The largest propagation step in the numerical simulations is $m=2500$, which can be reached in experiments owing to the strong interferometric robustness of the coupled fiber setup to environmental (i.e. thermal and mechanical) perturbations  \cite{r44}. In Fig.7(a) one has $\alpha=1/2$  and $V_0=0.02$, clearly corresponding to dynamical delocalization. In Fig.7(b) one has $\alpha=\alpha_c=1$ and $V_0=0.01$, corresponding to an oscillatory dynamics (Bloch oscillation regime). Finally, in Fig.7(c) one has $\alpha=3/2$ and $V_0=0.01$, corresponding to the asymptotic localization regime.

\section{Conclusions}
The localization, scattering and transport features of sliding potentials on a lattice are strongly dependent on the drift velocity owing to violation of the Galilean invariance for the discrete-space Schr\"odinger equation \cite{r19}. This leads to a variety of interesting effects that cannot be observed in systems described by a continuous-space   Schr\"odinger equation, such as the ability to make any scattering potential reflectionless when fast moving on the lattice \cite{r22,r23,r24} and washing out of Anderson localization and topological non-Hermitian phase transitions \cite{r26,r27}. Here we unraveled a new type of localization-delocalization phase transition, which is observed when a potential well with power law shape rapidly drifts on a lattice. For a power exponent $\alpha$ lower than the critical value $\alpha_c=1$, dynamical delocalization is observed, while for  $\alpha>\alpha_c$ asymptotic localization, corresponding to a frozen dynamics, is found. 
At the critical point $\alpha=\alpha_c$, where the discrete Schr\"odinger equation displays Galilean invariance, the system displays oscillatory dynamics, corresponding to Bloch oscillations.
As compared to common metal-insulator phase transitions in tight binding lattices with random or incommensurate disorder, where the phase transition arises when the strength of disorder is varied \cite{r30}, in our system there is not any disorder and the phase transition is not related to the amplitude of the potential, rather to the power exponent, and thus it is dubbed {\em algebraic} phase transition. Such a phase transition should be experimentally observable using synthetic photonic lattices, where engineered sliding potentials on the lattice can be readily implemented.
\medskip

% Acknowledgements
\medskip
\textbf{Acknowledgements} \par %delete if not applicable))
The author acknowledges the Spanish State Research Agency, through the Severo Ochoa
and Maria de Maeztu Program for Centers and Units of Excellence in R\&D (Grant No. MDM-2017-0711).

% References
\medskip

% Use the following code if you wish to generate your bibliography with BibTeX;
% replace the string "MSP-template" below with the name(s) of
% the BibTeX data base(s) you want to use.
% The resulting bibliography-output (the content of the .bbl file)
% must be pasted back into this file before submission.
% Please also include your BibTeX data base file(s) in your submission
% so that we can re-run BibTeX if necessary.
%
%\bibliographystyle{MSP}
%\bibliography{MSP-template}

% Figures/tables and captions
% Permission statements are required for all figures reproduced or adapted from previously published articles/sources. Please also ensure that all necessary permissions to reproduce images have been received
% Please remove these statements for original figures

% Please provide Biographies and photos for Essays, Feature Articles, Progress Reports, Reviews, and Perspectives for those authors who should be highlighted  
% These should be at most 100 words long
% For other article types this section can be removed
% Photographs should be 40mm broad and 50 mm high

%\begin{figure}
 % \includegraphics{bio-placeholder.jpg}
 % \caption*{Biography}
% \end{figure}

% Table of contents entry should be 50 - 60 words long
% Image should be 55 mm broad and 50 mm high or 110 mm broad and 20 mm high

 \appendix
 \renewcommand{\theequation}{A.\arabic{equation}}
\setcounter{equation}{0}
\small

\section{Wave dynamics in the fast moving limit: analytical study}
In this appendix we present an asymptotic analysis of the wave dynamics for the discrete Schr\"odinger equation (1) in the fast moving regime $v \gg \kappa$, which provides a simple physical explanation of the onset of the algebraic 
localization-delocalization phase transition.\\
Let us first recall that the largest speed of excitation propagation in the tight-binding lattice is given by $\sim 2 \kappa$, according to the Lieb-Robinson bound. For the initial excitation of the site $n=0$ of the lattice at time $t=0$, this means that at successive times excitation remains inside the cone $|n|< 2 \kappa t$. The discrete Schr\"odinger equation (1) with the potential given by Eq.(2) takes the explicit form
\begin{equation}
i \frac{d \psi_n}{dt}=- \kappa \left\{ \psi_{n+1}(t)+\psi_{n-1}(t) \right\}+V_0 (vt) ^{\alpha} \left|1-\frac{n}{vt} \right| ^{\alpha} \psi_n.
\end{equation}
For $|n| < 2 \kappa t$ and in the fast moving regime $v \gg \kappa$, we may set
\begin{equation}
\left|1-\frac{n}{vt} \right| ^{\alpha}  \simeq 1-\alpha \frac{n}{vt} 
\end{equation}
in Eq.(A.1), yielding
\begin{equation}
i \frac{d \psi_n}{dt}=- \kappa \left\{ \psi_{n+1}(t)+\psi_{n-1}(t) \right\}+V_0 (vt) ^{\alpha}  \left(1-\alpha \frac{n}{vt} \right) \psi_n.
\end{equation}
After the gauge transformation
\begin{equation}
\psi_n(t)= \phi_n(t) \exp \left( -i \frac{V_0 v^{\alpha} t^{\alpha+1}}{\alpha+1} \right) 
\end{equation}
from Eq.(A.3) one obtains
\begin{equation}
i \frac{d \phi_n}{dt}=- \kappa( \phi_{n+1}+\phi_{n-1})-F(t) n \phi_n
\end{equation}
 where we have set
 \begin{equation}
 F(t) \equiv \alpha V_0 (vt)^{\alpha-1}.
 \end{equation}
 Equations (A.5) describes the dynamics of a quantum particle on a tight-binding lattice under a time-dependence force $F(t)$, which can be solved analytically \cite{r48}. For initial single-site excitation $\psi_n(0)=\delta_{n,0}$, one obtains
 \begin{equation}
 | \psi_n(t)|^2= | \phi_n(t)|^2=J^2_{n} \left( 2 \kappa \sqrt{g(t)} \right)
 \end{equation}
 where $J_n$ is the Bessel function of first kind and of order $n$,
 \begin{equation}
 g(t) \equiv \left( \int_0^t d \xi \cos \eta(\xi) \right)^2+ \left( \int_0^t d \xi \sin \eta(\xi) \right)^2,
 \end{equation}
 and where we have set
 \begin{equation}
  \eta(t)=\int_0^t d \xi F(\xi)=V_0 v^{\alpha-1} t^{\alpha}.
 \end{equation}
 The evolution of the second order moment $\sigma^2(t)$ can be calculated analytically and reads \cite{r48}
 \begin{equation}
 \sigma^2(t)=2 \kappa^2 g(t).
 \end{equation}
 Clearly, a different scenario is found depending on whether $\alpha<\alpha_c$ or $\alpha> \alpha_c$, where $\alpha_c=1$. For $\alpha < \alpha_c$, the force $F(t)$ vanishes as $ t \rightarrow \infty$ and $\sigma^2(t)$ is unbounded, i.e. it secularly grows, as $t \rightarrow \infty$. The growth of $\sigma^2(t)$ with time $t$ is greatly dependent on the exponent $\alpha$, assuming a parabolic shape as $\alpha \rightarrow 0^+$ (ballistic transport) and  a periodic oscillatory shape (Bloch oscillation regime) as $\alpha \rightarrow 1^-$. This explains the qualitatively different behavior of delocalization observed in Fig.5 for $\alpha=1/4$ and $\alpha=1/2$. Conversely, for $\alpha> \alpha_c$ the force $F(t)$ diverges as $t \rightarrow \infty$, and $g(t)$ asymptotically reaches a stationary finite value as $t \rightarrow \infty$, indicating that the dynamics is asymptotically frozen. In particular, for $\alpha=2$, the force amplitude linearly increases with time, the integrals entering in Eq.(A.8) are the Fresnel integrals, and this problem was previously investigated in Refs.\cite{r28,r29} as an example of asymptotic localization. Finally, at the critical point $\alpha=\alpha_c=1$, the force $F(t)$ is constant in time and one obtains the Bloch oscillation dynamics.
 
 % \renewcommand{\theequation}{B.\arabic{equation}}
%\setcounter{equation}{0}

%\section{Energy spectrum on the semi-infinite lattice}
%The spectral problem of the two-particle non-Hermitian Hubbard model on a semi-infinite lattice is defined by Eqs.(5) with the boundary conditions (10) and (11). In the spirit of the Bethe Ansatz \cite{r3,GDV}, generalized to account for the imaginary 

\end{document}